\documentclass[12pt,preprint]{aastex}

\begin{document}

\title{The Fine Structure and Outskirts of DDO~154}

\author{G. Lyle Hoffman}
\affil{Dept. of Physics, Lafayette College, Easton, PA  18042; 
hoffmang@lafayette.edu}

\author{E.E. Salpeter}
\affil{Center for Radiophysics and Space Research, Cornell University, Ithaca, NY  14853; 
ees12@cornell.edu}

\and

\author{Nathan J. Carle\altaffilmark{1}}
\affil{Dept. of Physics, Lafayette College, Easton, PA  18042}
\altaffiltext{1}{Current address: Friend's Central School, 1101 City Ave., Wynnewood, PA  19096; ncarle@fcs.pvt.k12.pa.us}

\received{}
\accepted{}
\slugcomment{accepted for publication in AJ, Nov. 2001 issue}

\begin{abstract}

Mapping of the \ion{H}{1} disk of the isolated irregular galaxy DDO~154 with the C array of 
the Very Large Array and with the $3\farcm2$ upgraded Arecibo beam is presented.
Our results show a truncation (or temporary drop) of the \ion{H}{1} disk at a column density around 
$10^{19}$~atoms~${\rm cm}^{-2}$, consistent with theoretical expectations for the truncation produced by the extragalactic UV field.
We also detect a marginally significant levelling off of the \ion{H}{1} distribution along the continuation of the major axis at a column density near $2 \times 10^{18}$~atoms~${\rm cm}^{-2}$. 
The VLA results show that the gas beyond $\sim 6\arcmin$ in radius must be relatively smooth, with no structure larger in size than $\sim 300$ pc exhibiting a density contrast of a factor of 10 or more.
However, there is considerable few-hundred-parsec scale structure in the gas disk at smaller radii, even well outside the regions where there are visible stars.
Two prominent cavities well removed from any significant stellar populations are studied.
While the energies required for evacuation are consistent with those produced by multiple 
supernovae, there is no visible trace of stars within a kpc of the center of the larger 
cavity, and the smaller of the two cavities is centered just outside the 26.5 mag ${\rm 
arcsec}^{-2}$ $B$ isophote.
The velocity dispersion of the gas, measured within our 270 pc beam, is 7 to 8~km~${\rm 
s}^{-1}$ throughout the disk (to $6\arcmin$ radius).
This translates to a scaleheight of $\sim 700$ pc at the point where the rotation curve flattens, at a radius of $\sim 4.5$ kpc.
Velocity profiles are well fit by single gaussians at all points.

\end{abstract}


\section{Introduction}

The nearby irregular galaxy DDO~154 has received a great deal of attention for its  
extraordinarily extended \ion{H}{1} disk \citep{KB84,HLSFLR93,CP98} and its 
global dynamics, which seem to be dominated by dark matter at all radii \citep{CF88,CB89} but require some physics beyond that of the \citet{NFW96,NFW97} 
``universal'' halo \citep{BS97,GS99,GMGLQS00,MB88}.
Since the \ion{H}{1} disk extends so far beyond the visible stars while the galaxy is 
apparently isolated from all neighbors, in contrast to most other examples of \ion{H}{1} envelopes that extend far outside their optical components, DDO~154 would seem to be an excellent case in which to study the structure and thermodynamic state of a relatively pristine gas cloud.
There are two major questions:  (1)  How does the neutral gas disk terminate --- is it 
truncated abruptly by the extragalactic UV radiation field \citep{CS93,Ma93,DS94}, by UV photons leaking out from the central stellar disk \citep{BHFQ97}, or by some other process?
(2)  What is the fine structure and velocity dispersion of the outer gas disk --- is that gas, presumably well removed from energy input from star formation and supernovae, smoothly 
distributed or as filamentary as the inner parts of the Magellanic Clouds \citep{SSDSS99,KSDFSKM98} or well resolved spiral galaxies \citep{B97}?

To broach (1), we extended our Arecibo map of the outskirts of the galaxy, taking full 
advantage of the enhanced sensitivity and greatly reduced sidelobes of the upgraded telescope.
For (2), we undertook to improve, by a factor of three, the angular resolution of the mapping of the \ion{H}{1} envelope to assess the fine structure of the gas disk and in the hopes of finding sufficient column density to perform emission/absorption studies against one or more of the NVSS continuum sources in the outskirts of the SW quadrant of the \ion{H}{1} disk.
The average column density from the D array, DRAO and Arecibo mapping cited above would be too low, but if the outer envelope proved to be filamentary there would be a reasonable probability that such studies would be feasible.

In Sect. 2 our Arecibo study of the outskirts of the southernmost quadrant of the galaxy is presented.
We detail our VLA C array map of the galaxy in Sect. 3.
Sect. 4 is a discussion of the various theoretical points of interest, and
a summary and conclusions follow in Sect. 5.

\section{Arecibo Observations}

Our mapping of the outskirts of the southern quadrant of DDO~154 was conducted as a 
commissioning phase project at the upgraded Arecibo Observatory
\footnote{The Arecibo Observatory is part of the National Astronomy and Ionosphere Center, 
which is operated by Cornell University under a management agreement with the National Science 
Foundation.}
in July/August 1998.
We used the Gregorian feed system with the ``L narrow'' receiver in total power 
(position-switched) mode, with 6.10 kHz (about 1.3~km~${\rm s}^{-1}$) channel spacing.
Calibration was accomplished by observing several continuum sources from the VLA calibrator list, chosen to have small size compared to the $3\farcm2$ beam.
In addition, we reobserved several spiral galaxies for which we had high signal-to-noise pre-upgrade \ion{H}{1} measurements and which were known to be $\ll 3\farcm2$ in extent.

At the time of these observations, final focussing of the Gregorian feed system was still underway and the first sidelobe ring at 21cm was less uniform than in the final configuration and varied significantly with azimuth and zenith angle.
For a fuller characterization of the first sidelobe ring of the current instrument, see \citet{HPN+01}.
By scanning the beam across strong point sources at close to DDO~154's declination as they drifted across the field of view, we were able to determine that we could keep the center of the galaxy in the weakest part of the sidelobe ring (15dB or more down from the peak of the beam) as long as the point being observed lay between the center of the galaxy and the center of the dish at that moment.
Consequently, we planned our observations day by day to make sure that we kept that 
orientation, starting with points on the SE side of the galaxy as it entered the field of view and working around the rim to points on the SW side as the galaxy left the field of view.

The sidelobe contributions are not negligible, however, even though they are greatly reduced from those of the pre-upgrade circular feed.
The beam had a null ring (at least 20dB down) about $9\arcmin \times 7\farcm5$ in diameter, and the first sidelobe peaked in a ring about $11\farcm6 \times 10\farcm6$ in diameter with a maximum that varied from about $-16$dB to about $-13$dB; as discussed above, only the weaker part of the ring extended toward the center of the galaxy in our observations.
A $-15$dB uniform ring would have a total effective area 25-30\% of the main beam, so that one 
sextant of the ring would give about 4\% of the response of the main beam.
To correct for those sidelobe contributions, we interpolated between the points on our 
pre-upgrade flat feed map of the inner portions of the galaxy \citep{HLSFLR93} to estimate the flux that would fall in each sextant of the sidelobe ring, and then subtracted 4\% of that flux from the spectrum obtained at the position of the main beam.
The spectrum obtained at each observed position and the estimated sidelobe contribution to that spectrum are displayed in Figure 1.
The positions in relation to those mapped with the pre-upgrade flat feed in \citet{HLSFLR93} are shown in Fig. 2.

There is an additional worry:  scattering from the triangular platform and support cables is expected to produce far sidelobes of uncertain magnitude (but considerably less than the first sidelobe ring) which vary greatly with azimuth and zenith angle.
That expectation is consistent with our observation that the ability of the first sidelobe ring to account for the observed flux varies erratically from point to point, at least away from the extension of the major axis.
Far sidelobes seem to be a more reasonable explanation for that patchy low-level \ion{H}{1}, since the emission not accounted for by the first sidelobe has a velocity width comparable to that of spectra from points near the galaxy's center, not as narrow as the outer VLA C array spectra shown in the following section.
For the far sidelobes to account for the residual flux we see, the magnitude of the response for the portion of the far sidelobe which catches the center of the galaxy would have to be about $-25$dB for a wide range of distances from the center of the beam.
Nevertheless, in those cases where the remnant flux falls within the velocity bounds of the profiles obtained near the center of the galaxy, we conservatively attribute that flux entirely to incompletely removed sidelobes.
The only points with marginally significant emission that we cannot so attribute to sidelobes are those at the end of the warped major axis, labeled ``J'' and ``K,'' since that flux is seen at velocities higher than any from the central parts of the galaxy.
Taking, conservatively, the uncertainty in the corrected flux to be $1/3 \times 2 \times ( 50~{\rm km}~{\rm s}^{-1} ) \times ({\rm rms}~{\rm flux})$ added in quadrature to the uncertainty in the sidelobe contribution (approximately the square root of the number of contributing points times 20\% of the sidelobe flux), the sidelobe-corrected emission we detect at points J and K is marginally significant at levels of $2 \sigma$ and $3 \sigma$, respectively.
In Fig. 3 we show the sum of those two spectra.
Calculating uncertainties as above, the summed signal is significant at just over $3 \sigma$.

From the measured fluxes, after subtraction of the first sidelobe contributions, \ion{H}{1} column densities can be calculated.
The profiles obtained for points in this quadrant of the galaxy are approximately $50~{\rm km}~{\rm s}^{-1}$ broad, so in those cases where the sidelobe subtraction has left $< 2 \times ( 50~{\rm km}~{\rm s}^{-1} ) \times ({\rm rms}~{\rm flux} )$ we take that expression to be the effective upper limit on the corrected flux.
The fluxes and column densities, both as measured and corrected, are given in Table 1. 

The (warped) major axis of the galaxy follows points 1-4, J and K through the SW quadrant.
A second radial spoke, for comparison, can be taken to include points y, z, E and F.
The galaxy's center is at the point of convergence of the two spokes, marked by a filled circle.
The run of column density (corrected for first sidelobe contamination) vs. radius along those two spokes is shown in Fig. 4.
For the major axis, the center and points 1 through 4 are from our pre-upgrade flat-feed data as reported in \citet{HLSFLR93}.
Uncertainties in these points are of order 10\%.
The outer two points are from our new data (J and K) for the SW end of the major axis.
For these, at each radius we display two points connected by a vertical bar.
The upper point ignores sidelobe contamination and should be regarded as an absolute upper limit to the column density at that radius.
The lower point is corrected for sidelobe contamination as discussed above.
On these two particular spokes, the sidelobe-corrected fluxes exceed $2 \times ( 50~{\rm km}~{\rm s}^{-1} ) \times ({\rm rms}~{\rm flux} )$ at every point.
For the off-major axis, however, we cannot rule out incompletely removed sidelobes as the source of the flux as discussed above.
Therefore we are led to interpret these results as indicating a truncated edge to the neutral hydrogen disk at a radius of about 15\arcmin~ and a column density around $10^{19}~{\rm atoms}~{\rm cm}^{-2}$, with marginal evidence for the column density levelling out at $\sim 2 \times 10^{18}~{\rm atoms}~{\rm cm}^{-2}$ beyond that point on the major axis.
The uncertainty in the outermost column density is on the order of a factor of 2, however, and the conservative viewpoint would be that we have only an upper limit on the column density at point K.

\section{VLA C Array Mapping}

\ion{H}{1} spectral line mapping using 27 antennas in the C array of the Very Large Array
\footnote{The Very Large Array of the National Radio Astronomy Observatory is a facility of 
the National Science Foundation, operated under cooperative agreement by Associated 
Universities, Inc.}
was conducted on 24 Dec 1998.
Observational details are given in Table 2.
The pointing center was about 7\arcmin ~SW of the center of the galaxy, to give us greater sensitivity in the region where the \ion{H}{1} is most extended and where the NVSS sources are concentrated.
To obtain information about the velocity dispersion in the outer parts of the \ion{H}{1} envelope, we used 2.5~km~${\rm s}^{-1}$ velocity channels.
Online Hanning smoothing was employed, and calibration was accomplished using sources 1219+285 (B1950) and 3C286 from the VLA calibrator list.
The data were calibrated and editted using standard tasks in the Astronomical Image Processing System (Classic AIPS).
Continuum subtraction was done in the $uv$ plane via UVBAS, and maps were made and CLEANed using IMAGR with zero-spacing fluxes estimated from our Arecibo map.
Two cubes were made, one with robustness set equal to 0 (beam $18\arcsec\times 13\arcsec$) and one with robustness 1 (beam $18\arcsec\times 17\arcsec$).
After imaging, each data cube was corrected for the VLA primary beam.
For quantative results we have used the robustness 1 cube, since it is slightly more 
sensitive.
For moment analyses we blanked the cube using a mask made by smoothing the cube to a $60\arcsec \times 60\arcsec~$ beam, blanking everything more than 20 pixels outside the mask in each plane.

We detect \ion{H}{1} out to a radius of about 6\arcmin ~from the center of the galaxy, to a column density detection limit of $\sim 4 \times 10^{19}~{\rm atoms}~{\rm cm}^{-2}$ when the unblanked cube is smoothed to the typical linewidth ($19~{\rm km}~{\rm s}^{-1}$) of features within individual beams in the outer part of the \ion{H}{1} disk.
We conclude that the outermost gas is quite smooth on this scale, with no clumps that exceed $\sim 1 \times 10^{20}\,{\rm cm}^{-2}$ on the scale of our $18\arcsec\times 17\arcsec$ beam, or $280 \times 260$ pc assuming a 3.2 Mpc distance to the galaxy \citep{CB89}.
The more diffuse emission detected in our Arecibo map \citep{HLSFLR93} and in the D array plus DRAO map of \citet{CP98} has been resolved out by the lack of short baselines, and of course we can say nothing about structure on scales smaller than our beam size.
There are no indications of any region $18\arcsec$ or more in size with sufficient column density overlying any of the NVSS sources for absorption studies, unfortunately.
However, there is considerable structure in the \ion{H}{1} as far out as we can detect gas at this resolution, as shown in Fig. 5.
There the total \ion{H}{1} emission, integrated over velocity after blanking as described above, is shown.
It does not reach to our minimum detectable column density since blanking cannot be done perfectly; however, we have inspected a cube made by integrating over sets of 5 successive channels without blanking and found no features significantly outside the emission shown in Fig. 5.

\subsection{Two Prominent Cavities}

In particular, there are two prominent cavities centered approximately at $12^{h} 51^{m} 33^{s} + 27\arcdeg 24\arcmin 20\arcsec$(1950) and $12^{h} 51^{m} 44^{s} + 27\arcdeg 24\arcmin 50\arcsec$(1950).
In Fig. 6 each channel map is shown with a single contour indicating the outermost extent of the optical images from the Digitized Sky Survey (DSS).
The more easterly cavity lies on the very edge of the optical image, just beyond the 26.5 mag ${\rm arcsec}^{-2}$ $B$ isophote as shown in \citet{CB89}.
In the DSS image, there is a faint smudge that might be an \ion{H}{2} region near the center of the cavity.
The more westerly cavity, however, has no indication of any starlight whatsoever within it.
Neither can any H$\alpha$ emission be found associated with either cavity in the image acquired by \citet{HEB98}, reproduced in Fig. 7, nor in the deeper H$\alpha$ image of \citet{KS01}.

The diameters of the cavities are about 40\arcsec and 60\arcsec, respectively, or about 600 pc and 900 pc for our assumed 3.2 Mpc distance to DDO~154.
The velocity depths are about 20.6 and 33.6 km~${\rm s}^{-1}$ (corresponding to expansion velocities of order 10.3 and 16.8 km~${\rm s}^{-1}$).
However, as shown in Figs. 8 and 9 the spectra in the interiors of the cavities do not give clear indications of expanding walls at the front and rear; rather, the gas in the interior seems to partake of the same velocity field as the gas immediately adjacent to the cavity --- the neutral component in each cavity is simply less dense.
There are no indications of vorticity, although in the major axis-velocity map (Fig. 10) there is a dense ridge at the inner edge of the western cavity which clearly departs from uniform circular motion.

The velocity field contours from fitting gaussians to our C array mapping are shown in Fig. 5.
A position-velocity map, along the major axis and summed over the minor axis, is shown in Fig. 10.
The more easterly and more westerly holes fall at about 0 and $-80\arcsec$, respectively, on the position axis, and there is a dense ridge of material that deviates from circular motion just inside the westerly hole.
The rotation curve derived from the velocity field is shown in Fig. 11.
It does not differ significantly in the inner parts from the lower resolution mapping of \citet{CF88} or \citet{CP98}, so evidently beam-smearing \citep{SMT00,BRDB00} has not significantly affected the discussions of mass modeling cited above.
Further analysis and discussion of the rotation curve will be deferred to a forthcoming paper \citep{SHNSS01}.

The enhanced resolution in angle and velocity of our observations do allow us to determine the velocity dispersion of the gas.
Selected velocity profiles are displayed in Fig. 12; each is fit well by a single gaussian with no significant residuals.
A map of FWHM of gaussians fit to profiles at each pixel is presented in Fig. 13.
The corresponding dispersions are 7 to 8 ${\rm km}~{\rm s}^{-1}$ over most of the outer parts of the galaxy, rising to about 9 ${\rm km}~{\rm s}^{-1}$ in the central part (more or less coincident with the part that has formed stars).
The rising portion of the rotation curve would increase the velocity dispersion in the central regions (out to radius $50\arcsec$) of the galaxy by somewhat less than 2~km~${\rm s}^{-1}$, about what is observed, so it is not clear whether any part of the increase in the dispersion in the center can be attributed to heating by stars.
What is clear is that the dispersion in the outer parts, well outside the stellar portion, is still characteristic of the warm neutral medium (gas temperature of order $10^4$~K) rather than anything much colder.

The scaleheight of the gas in the outer parts of the disk can be estimated following \citet{PWBR92}.
We adopt $7.5~{\rm km}~{\rm s}^{-1}$ as a typical velocity dispersion at radii $4$--$6\arcmin~$.
The volume mass density in the disk at that radius is essentially all due to dark matter \citep{CP98}, which we take to be distributed spherically for convenience.
Then the mass density at radius $r$ is just

\begin{equation}
\rho (r) = \frac{1}{4 \pi G r^{2}} \frac{d}{dr} \left[ v(r)^{2} r \right]
\end{equation}

\noindent
where $v(r)$ is the circular velocity at radius $r$.
Where the rotation curve is flat, this reduces to

\begin{equation}
\rho (r) = \frac{v(r)^{2}}{4 \pi G r^{2}} .
\end{equation}

\noindent
In the case of DDO~154, the rotation curve goes flat at a radius of $\sim 4.5$~kpc \citep{CP98}, then declines gradually.
Taking the circular velocity to be $47~{\rm km}~{\rm s}^{-1}$ at that point, we find a volume density $\rho = 1.37 \times 10^{-25}~{\rm g}~{\rm cm}^{-3} = 0.002~{\rm M}_{\sun}~{\rm pc}^{-3}$ at that radius.
The scaleheight is 

\begin{equation}
h = \frac{\sigma}{[ 4 \pi G \rho ]^{1/2}}
\end{equation}

\noindent
\citep{Ke72, vdK81}, giving $h = 718$~pc for DDO~154 at a radius of 4.5~kpc ($4.\arcmin8$), well outside the stellar disk.

\section{Discussion}

\subsection{The outer envelope}

The most reasonable interpretation of the Arecibo results presented in Sect. 2 is that the \ion{H}{1} phase truncates rather abruptly, falling by nearly an order of magnitude in less than $3\farcm5$ (3.2 kpc) at a column density around $10^{19}~{\rm atoms}~{\rm cm}^{-2}$, 
with at best marginal indications of emission an order of magnitude lower continuing along the major axis.
As discussed in Sect. 2, the uncertainty in the column density at point J is of order a factor of 2, and the outermost point (K) should perhaps be regarded as an upper limit only.
While the suggested upturn in the rotation curve at large radii is intriguing, the data in hand do not warrant much further discussion of the marginally significant continuation of the major axis.
Still more Arecibo mapping, and a detailed study of the sidelobes of the upgraded Arecibo beam, is indicated.

That the neutral phase should abruptly decrease at such a column density is no surprise, since similar truncation has been observed in M33 \citep{CSS89,vG93}.
A theoretical explanation, in terms of photoionization by extragalactic UV, has been offered by \citet{CS93,Ma93,DS94}.
However, detection of H$\alpha$ flux around NGC~253 by \citet{BHFQ97} is claimed by those authors to require a more local photoionizing source, e.g., UV leaking out of the disk of the galaxy.
DDO~154, forming stars much more quiescently, is unlikely to be leaking enough H$\alpha$ to produce the truncation we claim to see so far outside the star-forming region.
A definitive test would be to compare the truncation along the line of nodes of the warped \ion{H}{1} disk with the direction perpendicular to that, since UV would be expected to escape mainly perpendicular to the disk of the galaxy and so could truncate the disk where a warp has lifted the edge of the disk out of the central plane but not along the line of nodes 
\citep{BHFQ97}.
Since we see, if anything, more abrupt truncation along the off-major axis spoke, which should be closer to the line of nodes, the data presented here suggests that the photoionizing flux is extragalactic in this case.

From our Arecibo data \citep[and Sect. 2]{HLSFLR93} and the VLA + DRAO map of \citet{CP98} we know that \ion{H}{1} remains above a few $\times 10^{19}~{\rm 
atoms}~{\rm cm}^{-2}$ for another 6 or 8\arcmin~ beyond that detected in the data of Sect. 3.
The outer gas must therefore be rather smooth, without structure that would amount to density contrasts of more than a factor of 2 or 3 on the scale of our 18\arcsec~ beam, since our C array detection threshold is a bit less than $10^{20}~{\rm atoms}~{\rm cm}^{-2}$ on that scale and much emission at ${\rm few} \times 10^{19}~{\rm atoms}~{\rm cm}^{-2}$ is evidently resolved out.
Structure such as we see in the inner 6\arcmin~, with density contrasts of a factor of 10 on 18\arcsec~ scales, does not continue into the outer envelope.
On the surface, this argues for the structure ultimately to be generated by stellar processes.
However, we argue in the following subsection that the connection between stars and \ion{H}{1} structure cannot be too intimate.

\subsection{Cavity generating mechanisms}

The cavities described in Sect. 3.2 are fairly typical of the holes found in Holmberg~II \citep{PWBR92}, IC 2574 \citep{WB99}, and DDO~47 \citep{WB01}, among others.
A similar cavity can be seen in the channel maps for NGC~2366 \citep{HEvW01}.
The required energies are consistent with those provided by multiple supernovae in a star formation region.
However, at least one of these cavities is well outside the region where {\em any} star formation has taken place.
Nothing is visible in the interior of the western cavity in B, R or H$\alpha$ images.
Similar claims have been made for many of the \ion{H}{1} holes in Holmberg~II \citep{RSWR99}, although those holes do not appear to be as far removed from the stellar regions of that galaxy.
Imaging in the UV might be more definitive \citep{SW00}, but to our knowledge no UV image with sufficient resolution has been published.
There is considerable resemblance to the giant \ion{H}{1} hole in NGC~6822 \citep{dBW00}, although DDO~154 gives no indication of tidal interaction such as that claimed for NGC~6822.

Another hole-making mechanism that has been discussed \citep[and references therein]{SFMK99} is a collision with a High Velocity Cloud.
We cannot rule that out in this case, although we otherwise have no evidence of there being HVC around DDO~154 and would not expect high impact velocities since the galaxy has low mass.
A third mechanism has been proposed recently by \citet{CTH00}) and 
supported by numerical hydrodynamical simulations \citep{WSK00}:  
cavities formed as gravitational instabilities in the process of spiral arm formation.
While there are no clear indications of spiral arms in either the optical or \ion{H}{1} images of DDO~154, a gravitational instability of this sort may still be the best explanation of these cavities. 
That would require, however, that the bulk of the mass of the galaxy be in the disk, 
distributed spatially as the \ion{H}{1} is distributed.
Other researchers have found that the best fit to the VLA + DRAO rotation 
curve has dark matter in a slightly flattened ($b/a = 0.6$) halo \citep{CP98} or in a disk as flat as the \ion{H}{1} disk \citep{HvAS01}, in each case with density proportional to that of the \ion{H}{1}.
Modified Newtonian Dynamics (MOND) has been claimed to give a good fit to the rotation curve \citep{MB88} with no dark matter at all.

\section{Conclusions and summary}

To study physical processes in the outermost \ion{H}{1} of a galaxy that is relatively 
undisturbed by tidal influences, we have pursued complementary mapping of DDO~154 with the Gregorian feed system at Arecibo Observatory and with the C array of the Very Large Array.
The $3\farcm2$ Arecibo beam gives us the most sensitive available information about the outermost edges of the \ion{H}{1} disk, and the C array mapping gives us information about the fine structure of the gas at somewhat smaller radii, but still well outside the stellar disk.

Despite uncertainty about the contribution of far sidelobes to our Arecibo spectra, we 
interpret our results to show a truncation of the \ion{H}{1} disk of DDO~154 at a column density around $10^{19}$~atoms~${\rm cm}^{-2}$, consistent with the truncations observed around the spirals M33 \citep{CSS89} and NGC~3198 \citep{vG93}.
This is consistent with theoretical expectations \citep{CS93,Ma93,DS94} for the truncation produced by the extragalactic UV field.
There are tantalizing hints that the column density along the continuation of the major axis might level off near $2 \times 10^{18}$~atoms~${\rm cm}^{-2}$ and at higher velocity than is seen in any other spectrum for the galaxy, but with only marginal significance. 
A detailed study of the sidelobes of the Gregorian feed system at Arecibo, and further mapping of DDO~154 with the Arecibo beam, will be required to verify these results.

The VLA results show that the gas beyond $\sim 6\arcmin$ in radius must be relatively smooth on scales of a few hundred pc or larger, with no structure at those scales exhibiting a density contrast of a factor of 10 or more.
We do not find sufficient column density against any background continuum sources to warrant absorption studies.
However, there is considerable structure in the gas disk at smaller radii, even well outside the regions where there are visible stars.
Two prominent cavities well removed from any significant stellar populations were noted, with a pronounced overdensity between them.
While the energies required for evacuation are consistent with those produced by multiple supernovae, there is no visible trace of the stars which could have engendered those supernovae within a kpc of the center of the larger cavity.
The smaller of the two cavities is centered just outside the 26.5 mag ${\rm arcsec}^{-2}$ $B$ isophote.
Other hole-making mechanisms were discussed in Sect. 4.2, but none presents itself as the obvious choice.

The velocity dispersion of the gas, measured within our $18\arcsec$ (270 pc) beam, is 7 to 8~km~${\rm s}^{-1}$ throughout the disk (to $6\arcmin$ radius), with a slight enhancement at the galaxy's center (which could be produced either by increased turbulence associated with star formation or by the rising part of the rotation curve) and a slight diminishment in the center of the larger cavity.
Velocity profiles are well fit by single gaussians at all points.
This translates to a disk scaleheight of 718~pc at the radius where the rotation curve turns flat, about 4.5~kpc.

Detailed discussion of the rotation curve will follow in a separate paper \citep{SHNSS01}.

\acknowledgments
D. Hunter kindly made available to us her H$\alpha$ image of DDO~154.
We thank J. van Gorkom for guidance on observing procedures and suggestions on analysis, and R. Swaters and R. Braun for valuable discussions.
Partial support was provided by the Crafoord Fund at Cornell University and by a faculty research grant from Lafayette College.
The Digitized Sky Surveys were produced at the Space Telescope Science Institute under U.S. Government grant NAG W-2166. 
The images of these surveys are based on photographic data obtained using the Oschin Schmidt Telescope, which is operated by the California Institute of Technology and Palomar Observatory on Palomar Mountain. 
The plates were processed into the present compressed digital form with the permission of that institution.
The Second Palomar Observatory Sky Survey (POSS-II) was made by the California Institute of Technology with funds from the National Science Foundation, the National Geographic Society, the Sloan Foundation, the Samuel Oschin Foundation, and the Eastman Kodak Corporation.
This research has made use of the NASA/IPAC Extragalactic Database (NED) which is 
operated by the Jet Propulsion Laboratory, California Institute of Technology, under 
contract with the National Aeronautics and Space Administration.

\clearpage

\begin{deluxetable}{c c c c c c c c c c}
\tabletypesize{\footnotesize}
\rotate
\tablewidth{0pt}
\tableheadfrac{0.3}
\tablecaption{Arecibo Results}
\tablecolumns{10}
\tablehead{
\colhead{Label} & \colhead{Radius} & \colhead{P.A.} & \colhead{RA(1950)} & \colhead{Dec(1950)} & \colhead{Raw Flux} & \colhead{rms} & \colhead{Corr. Flux} & \colhead{Raw $N_{HI}$} & \colhead{Corr. $N_{HI}$} \\
\colhead{} & \colhead{\arcmin} & \colhead{\arcdeg} & \colhead{hhmmss.s} & \colhead{ddmmss} & \colhead{mJy km ${\rm s}^{-1}$} & \colhead{mJy} & \colhead{mJy km ${\rm s}^{-1}$} & \colhead{$10^{17}~{\rm cm}^{-2}$} & \colhead{$10^{17}~{\rm cm}^{-2}$} }
\startdata
A & 10.4 & 123 & 125218.7 & 271948 & 229 & 0.61 & $<61$ & 64 & $<17$ \\
B & 13.5 & 133 & 125223.8 & 271604 & 122 & 0.65 & $<106$ & 34 & $<29$ \\
C & 10.3 & 144 & 125206.7 & 271657 & 244 & 0.72 & $<72$ & 68 & $<20$ \\
D & 14.0 & 149 & 125211.8 & 271313 & 136 & 0.58 & $<112$ & 38 & $<31$ \\
E & 11.7 & 163 & 125154.7 & 271403 & 232 & 0.75 & $<75$ & 65 & $<21$ \\
F & 15.6 & 163 & 125159.8 & 271022 & 101 & 0.46 & $<84$ & 28 & $<23$ \\
G & 17.9 & 174 & 125147.9 & 270731 & $~$89 & 0.50 & $<82$ & 25 & $<23$ \\
H & 14.1 & 177 & 125142.7 & 271112 & 227 & 0.77 & $<120$ & 63 & $<33$ \\
I & 17.1 & 187 & 125130.7 & 270821 & $~$62 & 0.40 & $<40$ & 17 & $<11$ \\
J & 17.0 & 200 & 125113.6 & 270917 & 140 & 0.67 & $~$71 & 39 & $~$20 \\
K & 20.6 & 204 & 125101.7 & 270626 & $~$85 & 0.68 & $~$72 & 23 & $~$20 \\
L & 17.9 & 212 & 125056.4 & 271010 & $~$60 & 0.45 & $<45$ & 17 & $<13$ \\
M & 15.6 & 223 & 125051.3 & 271354 & 142 & 0.54 & $<54$ & 39 & $<15$ \\
N & 14.1 & 237 & 125046.3 & 271731 & 123 & 0.80 & $<80$ & 34 & $<22$ \\
\enddata
\end{deluxetable}

\begin{deluxetable}{l l}
\tabletypesize{\small}
\tablewidth{0pt}
\tablecaption{Very Large Array Observations}
\tablehead{}
\startdata
Date & 1998 Dec 24 \\
Pointing Center R.A.(1950) & 12:51:30.0 \\
Pointing Center Dec.(1950) & 27:20:00 \\
Heliocentric velocity & 375 km ${\rm s}^{-1}$ \\
Array & C \\
Channels & 128 \\
Channel separation & 2.6 km ${\rm s}^{-1}$ \\
Time on source & 609 min \\
Beam & $18 \times 13$ arcsec \\
Noise rms per channel & 0.76 mJy ${\rm beam}^{-1}$ \\
\enddata
\end{deluxetable}

\begin{figure}
\figurenum{1a}
\caption{
Spectra (solid curves) labelled by radius in arcmin and position angle in degrees.
The thin dashed curve is the estimated sidelobe contribution to each spectrum.
Labels from Table 1 are given in the upper right hand corner of each plot.}
\end{figure}

\begin{figure}
\figurenum{1b}
\caption{
{\it cont.}}
\end{figure}

\begin{figure}
\figurenum{1c}
\caption{
{\it cont.}}
\end{figure}

\begin{figure}
\figurenum{1d}
\caption{
{\it cont.}}
\end{figure}

\begin{figure}
\figurenum{1e}
\caption{
{\it cont.}}
\end{figure}

\begin{figure}
\figurenum{1f}
\caption{
{\it cont.}}
\end{figure}

\begin{figure}
\figurenum{1g}
\caption{
{\it cont.}}
\end{figure}

\clearpage

\begin{figure}
\figurenum{2}
\plotone{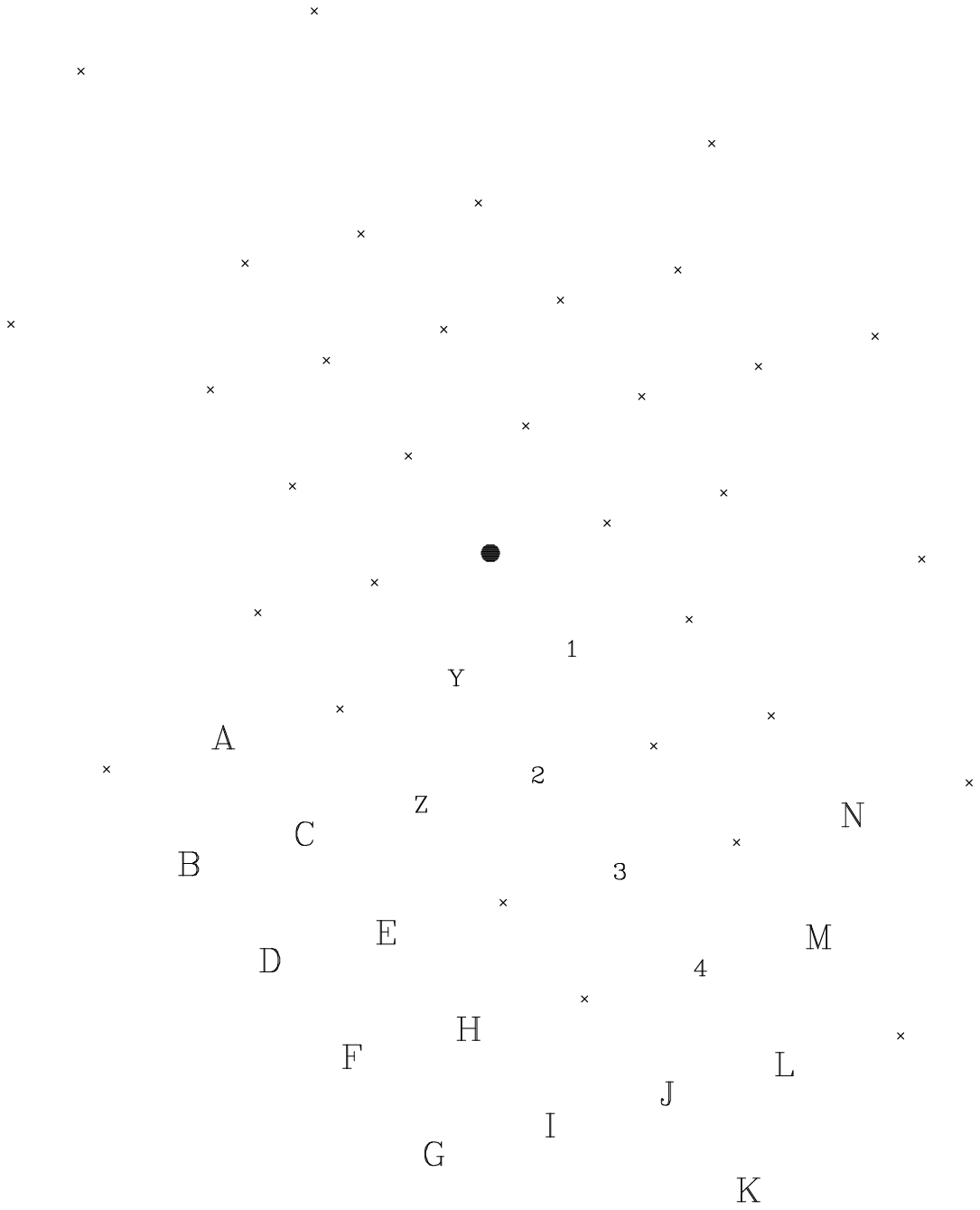}
\caption{
Map of positions on DDO~154 observed with the Arecibo $3\farcm2$ beam.
The newly observed points are indicated by their labels (large letters A-N) from Table 1.
All other points were reported in \citet{HLSFLR93}.
The galaxy's center is marked by a solid circle.
Points 1-4, J and K trace the warped major axis, approximately; points y, z, E and F 
constitute another spoke for later reference.
}
\end{figure}

\begin{figure}
\figurenum{3a}
\plottwo{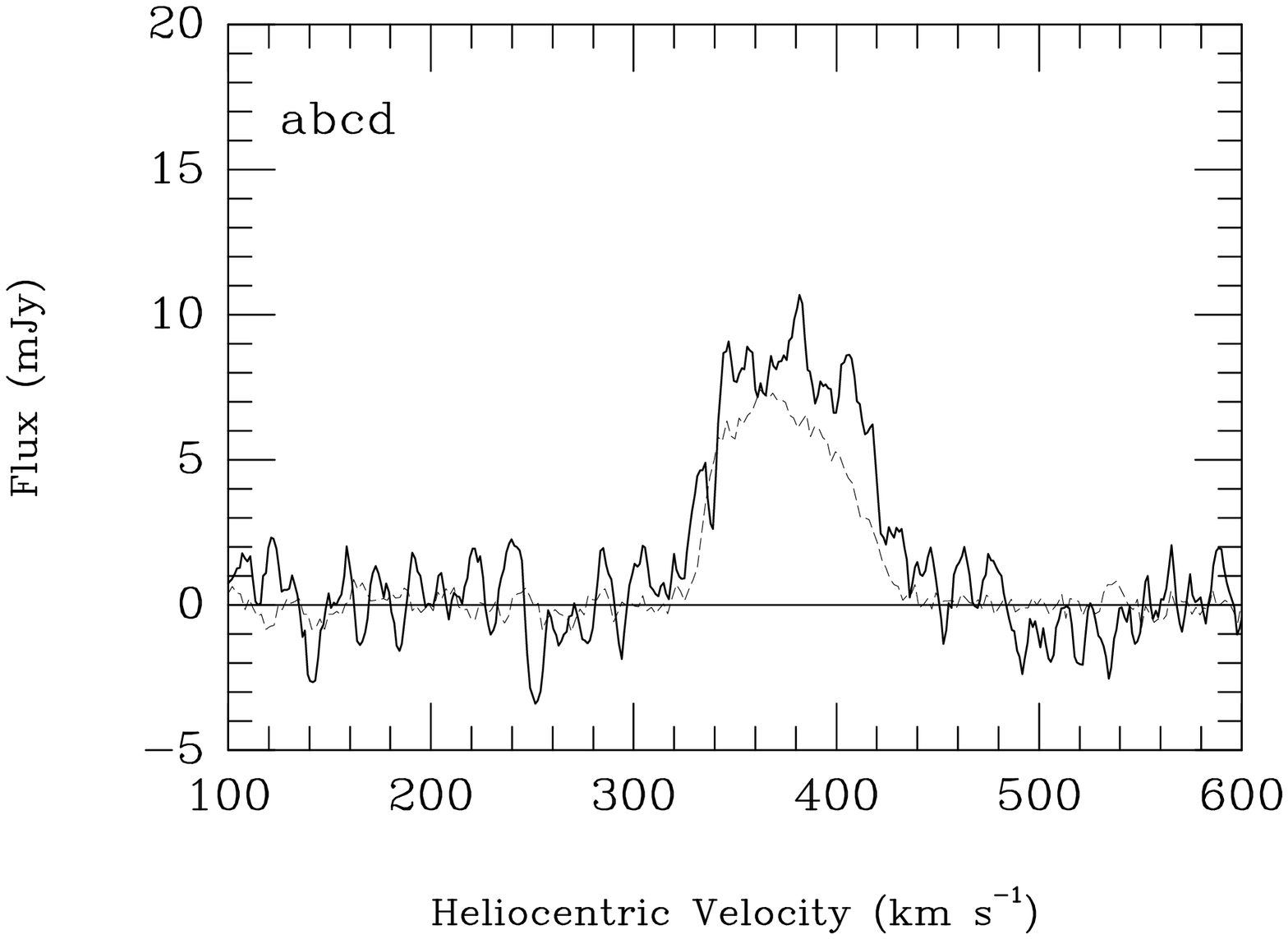}{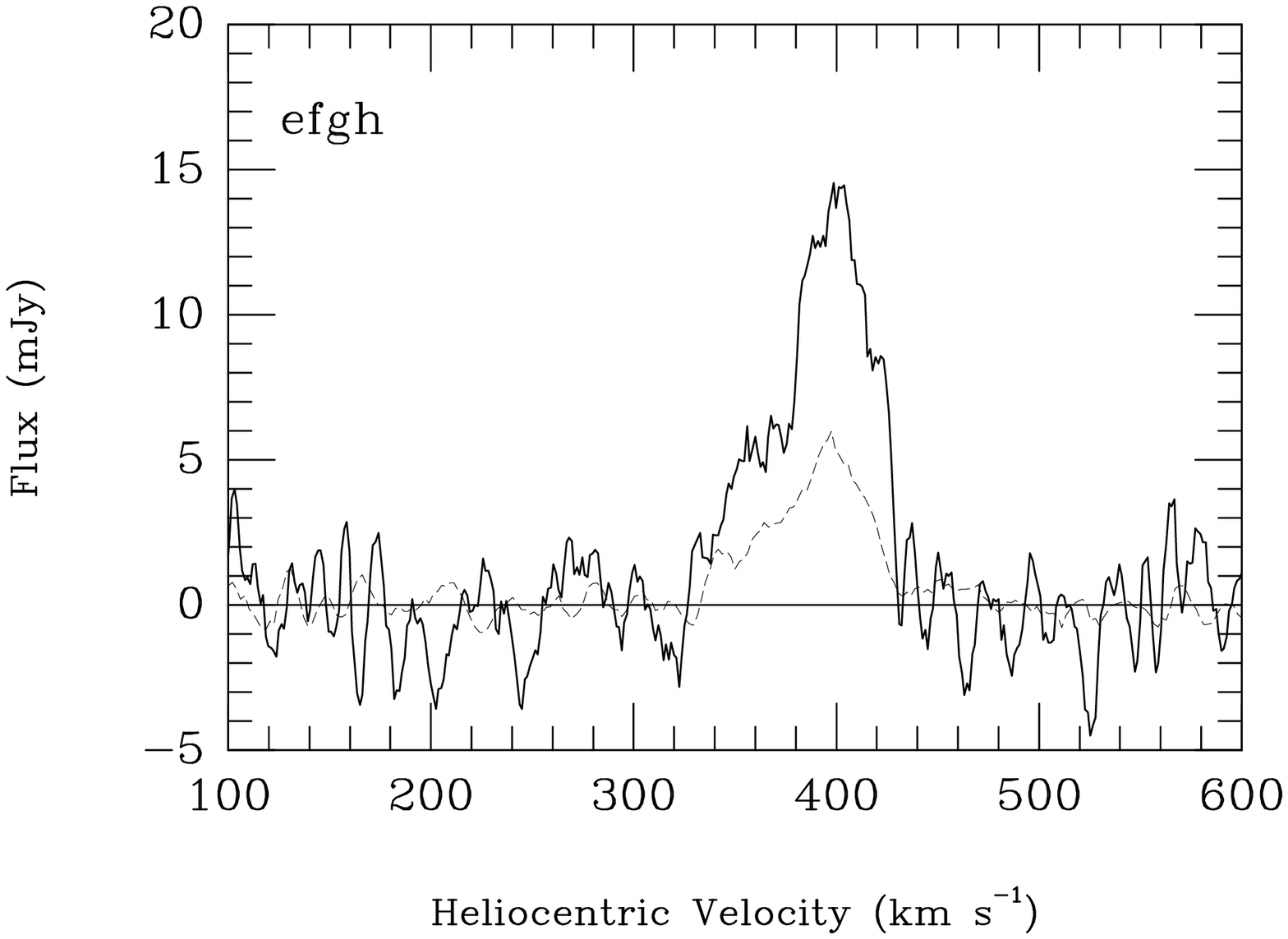}
\caption{
Sum of spectra (solid curve) from sets of adjacent points.
The thin dashed curve is the estimated summed first sidelobe contribution in each case.
Each panel is labelled by the points that contribute to the sum.}
\end{figure}

\begin{figure}
\figurenum{3b}
\plottwo{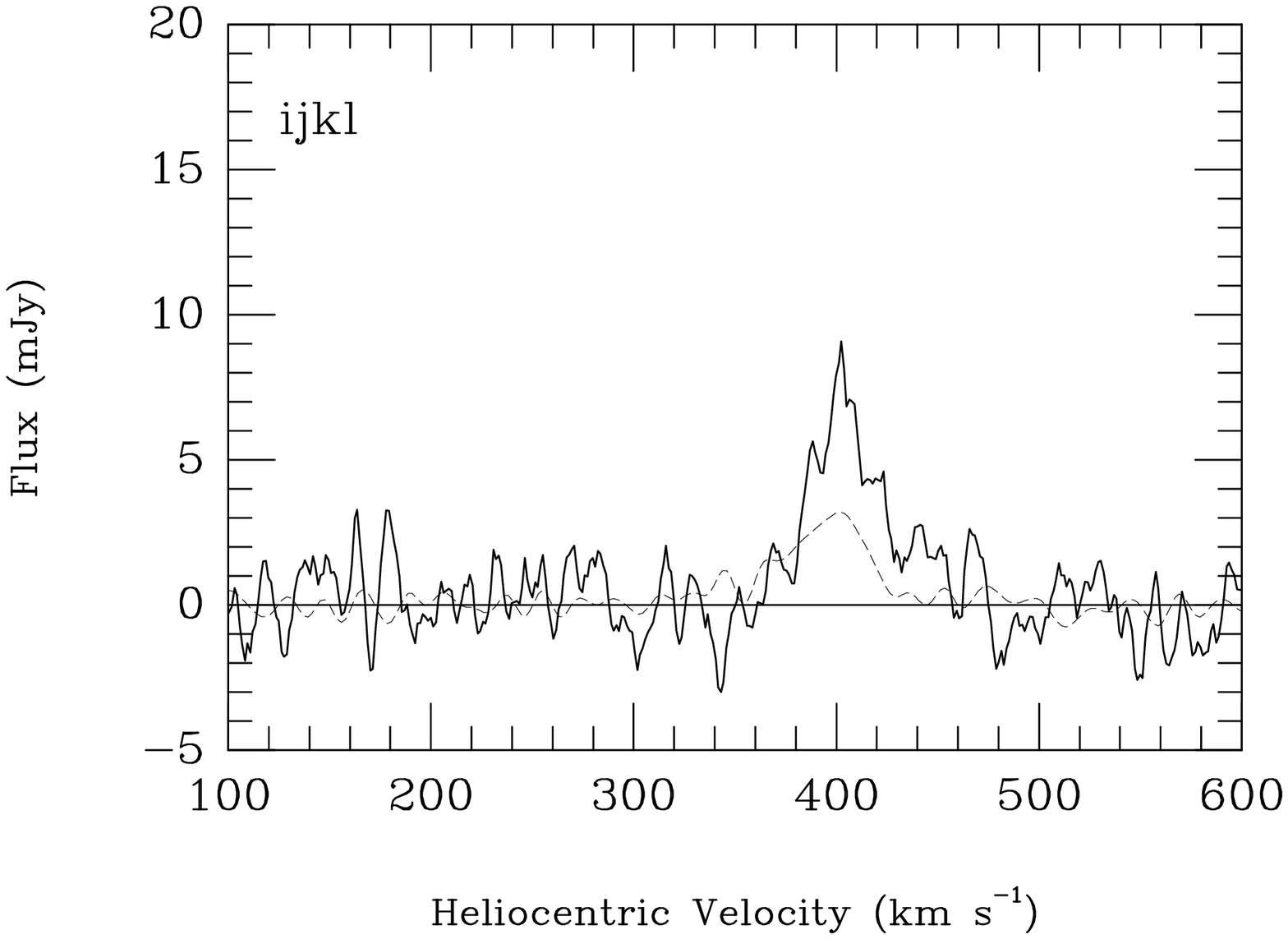}{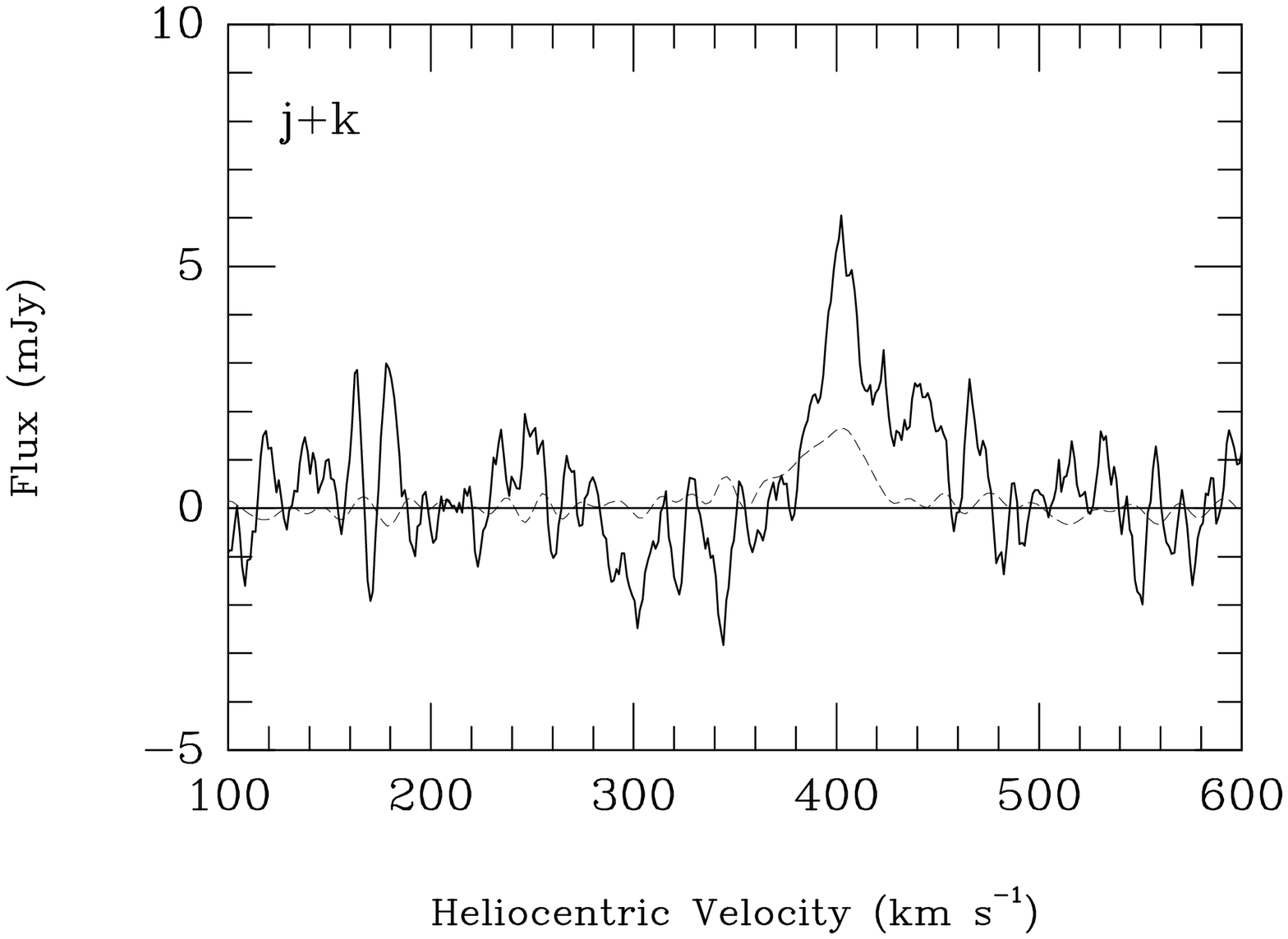}
\caption{
{\it cont.}}
\end{figure}

\begin{figure}
\figurenum{4}
\plotone{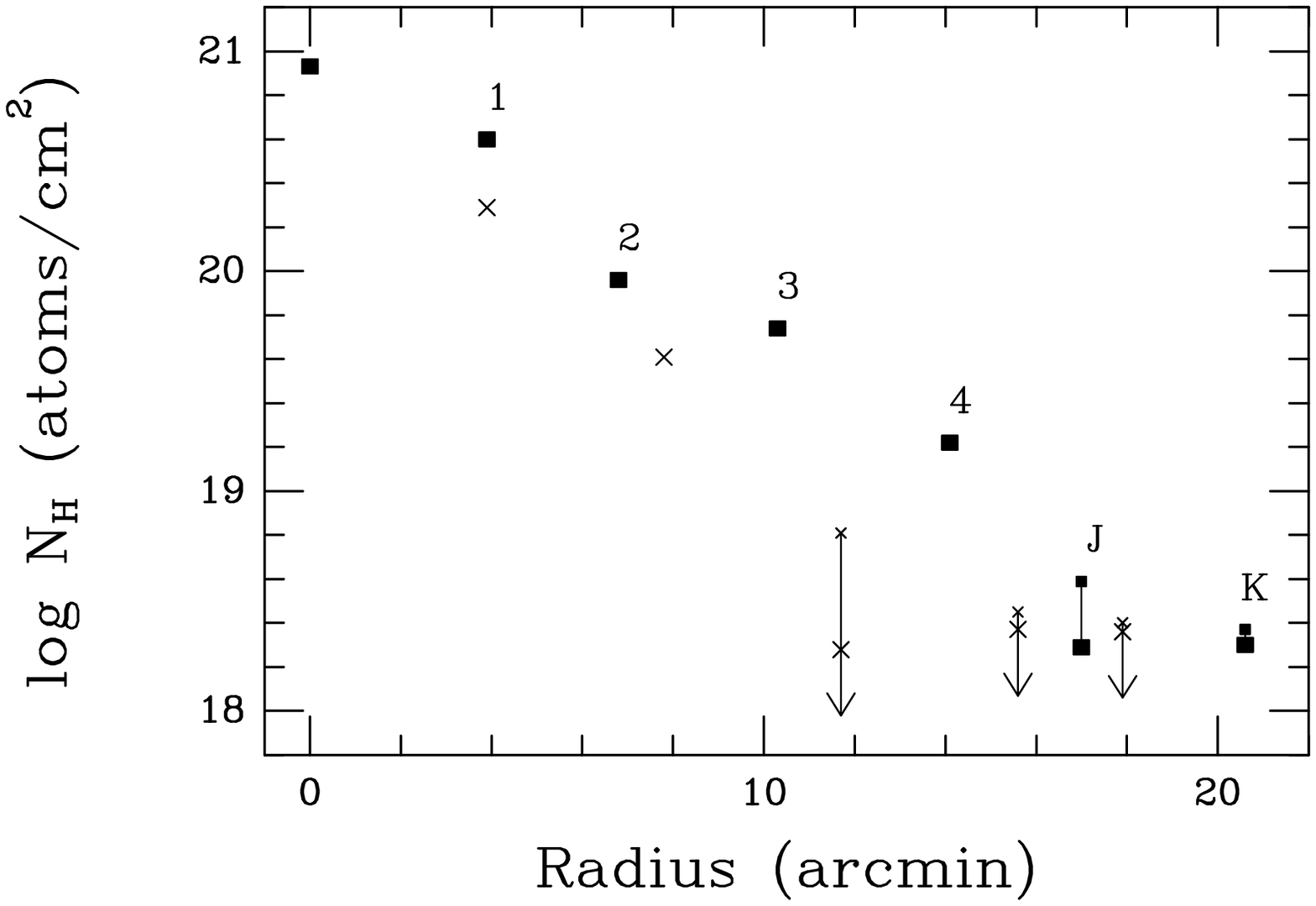}
\caption{
Column density as a function of radius along the major axis, following the warp (solid 
squares, from the galaxy's center and points 1-4, J and K in Fig. 2) and along a second spoke 
(exes, from points y, z, E and F in Fig. 2).
The newly acquired beam positions (post-upgrade) have two points connected by a vertical bar; 
the upper point in each case assumes no sidelobe contribution while the lower point has been 
corrected for a uniform $-15$dB sidelobe ring.
Spectrum J corresponds to the pair of solid squares at radius $17.
\farcm0$, spectrum K to the 
outermost pair of solid squares.
There are likely to be far sidelobe contributions to the off-major axis positions (exes) as 
discussed in the text, so we interpret the lower points at those positions to be upper limits 
as well, indicated by the downward arrows.
For the inner points (center, 1-3 and the two pre-upgrade points on the second spoke), uncertainties in column density are of order 10\%.
}
\end{figure}

\begin{figure}
\figurenum{5}
\caption{
Grey scale representation of the integrated hydrogen emission, in mJy-km~${\rm s}^{-1}$/B, 
from DDO~154, with velocity field contours (from fitting gaussians) superimposed.  
The contours range from 315 km~${\rm s}^{-1}$ on the NE end to 425 km~${\rm s}^{-1}$ in the 
SW, in steps of 10 km~${\rm s}^{-1}$.
The minimum displayed column density is about $1.3 \times 10^{20}~{\rm atoms}~{\rm cm}^{-2}$.
Arecibo beam positions along the warped major axis are marked with crosses, with a circle for the center.
The labels superimposed on the outer crosses correspond to those in Fig. 2.}
\end{figure}

\clearpage

\begin{figure}
\figurenum{6a}
\caption{
Grey scale images of neutral hydrogen emission in each velocity channel with an outer contour 
from a smoothed optical image from the Digitized Sky Survey superimposed.
The contour corresponds approximately to $26~{\rm mag}~{\rm arcsec}^{-2}$.
The gray scale is in units of mJy/Beam.}
\end{figure}

\begin{figure}
\figurenum{6b}
\caption{
{\it cont.}}
\end{figure}

\begin{figure}
\figurenum{6c}
\caption{
{\it cont.}}
\end{figure}

\begin{figure}
\figurenum{6d}
\caption{
{\it cont.}}
\end{figure}

\begin{figure}
\figurenum{6e}
\caption{
{\it cont.}}
\end{figure}

\clearpage

\begin{figure}
\figurenum{7}
\caption{
Neutral hydrogen contours from the total \ion{H}{1} map compared to a grey-scale representation of the H$\alpha$ emission.
In the left-hand panel, only the 250 mJy km~${\rm s}^{-1}\;{\rm Beam}^{-1}$ contour is shown to outline the two prominent cavities.
On the right, only higher contours (400, 500 and 600 mJy km~${\rm s}^{-1}\;{\rm Beam}^{-1}$) are shown to highlight the relationship between the H$\alpha$ and \ion{H}{1} peaks.
The prominent H$\alpha$-emitting cloud just west of the larger \ion{H}{1} cavity appears to be associated with a foreground star.
}
\end{figure}

\begin{figure}
\figurenum{8}
\plotone{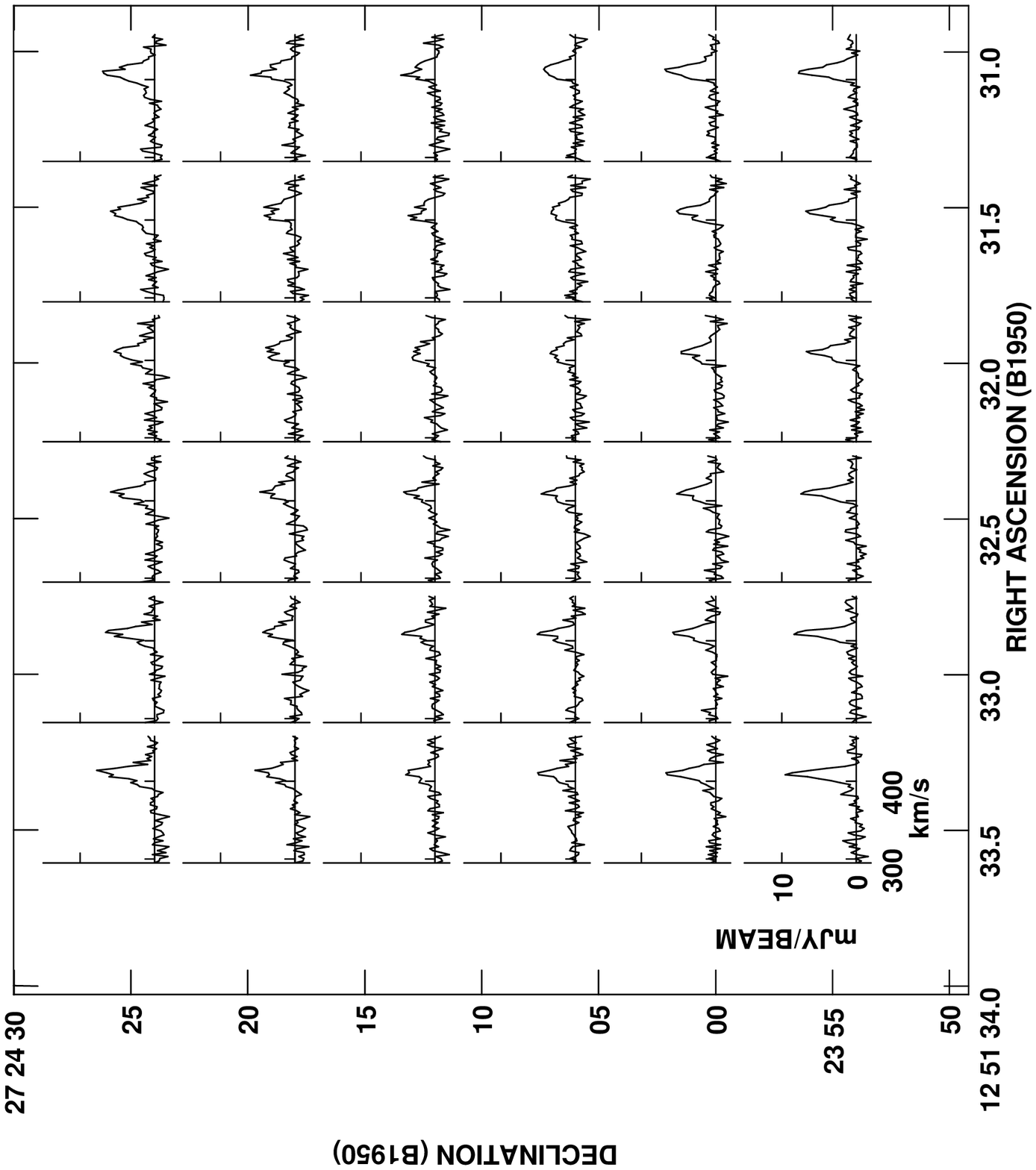}
\caption{
Velocity profiles across the more westerly cavity.
A spectrum is plotted for every second pixel.}
\end{figure}

\begin{figure}
\figurenum{9}
\plotone{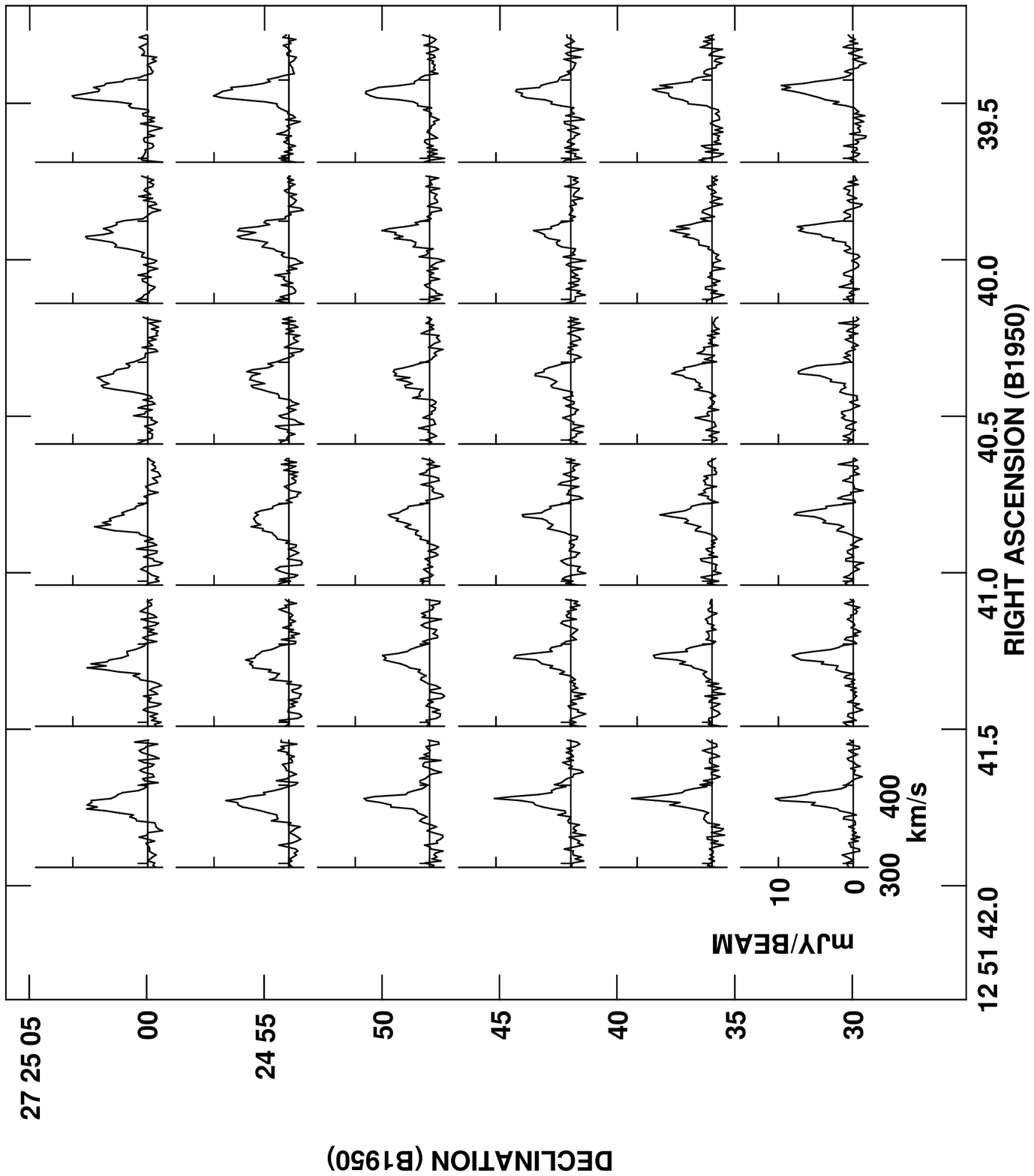}
\caption{
Velocity profiles across the more easterly cavity.
A spectrum is plotted for every second pixel.}
\end{figure}

\begin{figure}
\figurenum{10}
\caption{
Position-velocity map, aligned along the central part of the major axis and summed over the 
minor axis.
Grey scale units are mJy/Beam.
Coordinates are relative to the center of the galaxy.
The more western cavity falls at about $-80\arcsec$ on the position axis.}
\end{figure}

\clearpage

\begin{figure}
\figurenum{11a}
\caption{
Rotation curves for the galaxy as a whole (circles), and for the receding (upward pointing 
triangles) and approaching (downward pointing triangles) sides of DDO~154, determined by 
fitting the velocity field within consecutive annuli.
}
\end{figure}

\begin{figure}
\figurenum{11b}
\caption{
Inclinations (left panel) and position angles (right panel) of successive annuli for the 
galaxy as a whole (circles), and for the receding (upward pointing triangles) and approaching 
(downward pointing triangles) sides of DDO~154, determined as for Fig. 11a.
}
\end{figure}

\clearpage

\begin{figure}
\figurenum{12}
\caption{
Velocity profiles at selected points across DDO~154.}
\end{figure}

\begin{figure}
\figurenum{13}
\caption{
Grey scale representation of the velocity dispersion (FWHM of fitted gaussians) map of 
DDO~154.
The grey scale indicates FWHM in km~${\rm s}^{-1}$, ranging from 0 to 40~km~${\rm s}^{-1}$.
There may be some augmentation of the velocity dispersion by the gradient in the rotation 
curve in the central 50$\arcsec$~ of the galaxy, but the outer parts of the velocity 
dispersion field should not be significantly affected by such gradients.}
\end{figure}

\end{document}